# Quantum Key Recycling with Optimal Key Recycling Rate based on Error Rate

Yu-Chin Lu[0000-0001-5982-5611], Chia-Wei Tsai[0000-0002-8075-6558], and Tzonelih Hwang[0000-0003-3881-1362]*

*Abstract* —We propose a new Quantum Key Recycling (QKR) protocol, which can tolerate the noise in the quantum channel. Our QKR protocol recycles the used keys according to the error rate. The key recycling rate of the pre-shared keys in our QKR protocol is optimized depending on the real error rate in the quantum channel. And our QKR protocol has higher efficiency than the exiting QKR protocol with error-tolerance. The security proof shows the security of the recycled keys is universal composable.

## I. Introduction

AN important issue in quantum cryptography is key recycling. Taking One-time Pad (OTP) in classical cryptography as an example, if a sender Alice uses $k_i$ to encrypt a message $m_i$ (i.e., $c_i = m_i \oplus k_i$ ) and sends the ciphertext $c_i$ to the receiver Bob, can Alice recycle the used $k_i$ in the subsequent communications? Obviously, the used key cannot be recycled securely if an adversary, Eve, owns a copy of the ciphertext $c_i$ and the message $m_i$ pair (Eve can perform the known-plaintext attack in this situation). But if the cipher is encoded into the quantum state, it is possible to recycle the used key securely, even if Eve has unlimited computation power. This is because a quantum cipher can detect the existence of eavesdropping by using the quantum uncertainty principle.

Quantum Key Recycling (QKR) is first proposed in 1982 [1]. In a QKR protocol, Alice and Bob pre-share some keys firstly, Alice uses the pre-shared keys to encrypt the message and the Message Authentication Code (MAC) of that message, and then encode the cipher into quantum states. If Eve wants to get the information about the cipher, she will disturb the quantum states and affect the MAC. After Bob receives the qubits, he can decode the cipher and check the MAC. If the check passes, he can ensure that this communication is not eavesdropped by Eve, i.e., the cipher is not intercepted or copied by Eve. Because the cipher has not been known by Eve, the used keys can be recycled even the keys are used to produce quantum OTP. The studies [2, 3] proved that this idea is secure and the recycled keys have composable security.

The study [2] also indicates a new question: how to add the noise-tolerance property to the QKR protocol? The above-mentioned QKR protocols can only transmit the message via an ideal (noiseless) channel because the integrity of MAC will be broken and then the protocol will fail if any qubit is disturbed by the channel's noise. Since the assumption of the noiseless channel is not practical, this issue influences the feasibility of the QKR protocols in practical applications. [2] provided a solution to this question, but it can only tolerate a few noise errors. Therefore, other studies [4-6] tried to increase the noise-tolerance level by adding Error Correcting Code (ECC) such as an RS code [7] or LDPC code [8]. After Bob receives the qubits and obtains the cipher, he will use the ECC to correct the noise errors first and then check the MAC. These studies [4-6] need to predict the error rate in the quantum channel. And no matter there are how many real errors in the quantum channel, they need to consume some keys according to their prediction to keeping the recycled keys secure. This makes their protocol cannot recycle all used keys even when the quantum channel is actually noiseless[1].

In this paper, we propose a new QKR protocol for the noisy quantum channel. Our QKR protocol can recycle the used pre-shared keys according to the real error rate in the quantum channel. The core idea of our research is to take the number of errors that are corrected by ECC into account when recycling keys. Our QKR protocol uses single qubits with conjugate coding and doesn't need any classical communication. We eliminate the classical channel by delaying recycling some of the keys (these keys are recycled in the next round of the protocol). And our QKR protocol can be used to send a secret message with a quantum channel that has a very high error rate (The existing QKR protocols with conjugate coding [4, 6] can only run on a channel which its error rate is less than 0.11). We also give a security analysis of the recycled keys and optimize the key recycling rate (the length of the recycled keys / the length of all the keys used to run the protocol). Our QKR protocol can recycle all the keys when no error exists and can recycle keys safely and efficiently if the real error rate is less than or equal to the predicted error rate.

The security of the recycled keys is analyzed based on information theory. The security of the recycled keys in our

Manuscript received May 31, 2020. This research was partially supported by the Ministry of Science and Technology, Taiwan, R.O.C. (Grant Nos. MOST 107-2218-E-143-002-MY2, and MOST 107-2627-E-006-001 and 108-2221-E-006 -107)

Yu-Chin Lu is with the Department of Computer Science and Information Engineering, National Cheng Kung University, No. 1, University Rd., Tainan City, 70101, Taiwan, R.O.C (e-mail: m82617m@gmail.com).

Chia-Wei Tsai is with Department of Computer Science and Information Engineering, National Taitung University, No. 369, Sec. 2, University Rd., Taitung City, 95092, Taiwan, R.O.C. (e-mail: cwtsai@nttu.edu.tw).

Tzonelih Hwang is the corresponding author. He is with Department of Computer Science and Information Engineering, National Cheng Kung University, No. 1, University Rd., Tainan City, 70101, Taiwan, R.O.C. (e-mail: hwangtl@csie.ncku.edu.tw).

[1] A QKR-like protocol called unclonable encryption [9] Also abandons some keys to erase the information leaked to Eve. And the unclonable encryption protocol unable to recycle all the keys even when there is no error, too.

QKR protocol is universal composable [10]. The distance between the recycled keys and the ideal keys is limited by a specific security parameter $\varepsilon$. This means that an adversary only has an $\varepsilon$ chance to distinguish the difference between our QKR protocol and an ideal one.

## II. PRELIMINARIES

Most of the notations will be explained when they first appear in this paper, but we describe a few notations here for convenience: the notation $|X|$ denotes the number of possible values of a variable $X$ and we define $|X|_l = \log_2 |X|$.

### A. Information entropy

The concept of information entropy is introduced by Shannon in 1948 [11]. It gives a way to measure the randomness of a message. We denote the information entropy of a random variable $X$ in bits by

$$H(X) = -\sum_x p(x) \log_2 p(x),$$

where $p(x)$ is the probability of possible values $x$. $H(X)$ achieves the maximum when all possible values of $X$ have the uniform probability and will be 0 when the value of $X$ is definite. This entropy is also called the "Shannon entropy". The meaning of the Shannon entropy of a random variable $X$ is the uncertainty of the variable $X$.

We denote the conditional entropy of a random variable $X$ when given another variable $Y$ by

$$H(X|Y) = -\sum_{x,y} p(x,y) \log_2 \frac{p(x,y)}{p(y)}.$$

$H(X|Y) = 0$ if and only if the variable $X$ is completely determined by $Y$. Conversely, $H(X|Y) = H(X)$ if and only if $X$ is completely independent of $Y$.

Using the definition above, we can denote the mutual information of two variables $X$ and $Y$ by

$$I(X;Y) = H(X) - H(X|Y) \\ = H(Y) - H(Y|X).$$

The meaning of mutual information is the uncertainty of a variable when the other variable is known. It is the opposite of conditional entropy. When two variables are completely independent, the mutual information will be 0. And the mutual information will achieve maximum when one variable can completely determine the other one.

### B. The density operator and the von Neumann entropy

We denote the density operator of a quantum state $Q = |\psi\rangle$ as $\rho_Q = |\psi\rangle\langle\psi|$. We can use the density operator of the quantum state $Q$ to calculate the von Neumann entropy of $Q$. The von Neumann entropy is very similar to the Shannon entropy. They both are used to measure uncertainty. But the von Neumann entropy is used to measure the uncertainty of a quantum state instead of a variable. We denote the von Neumann entropy of a quantum state $Q$ by

$$S(Q) = -\text{tr}(\rho_Q \log_2 \rho_Q),$$

where tr denotes trace. And if we write $\rho_Q$ in terms of its eigenvectors $|1\rangle, \ldots, |j\rangle$ as $\rho_Q = \sum_j \eta_j |j\rangle\langle j|$, then the von Neumann entropy is [12]

$$S(Q) = -\sum_j \eta_j \log_2 \eta_j.$$

It is worth mentioning that the von Neumann entropy is not used to measure the measurement uncertainty of a quantum state but is used to measure the uncertainty of a mixed state. The von Neumann entropy is not affected by the measurement operators (or called measurement basis) used to measure the state. For example: If we use the z-basis to measure the quantum state $|+\rangle = 1/\sqrt{2}(|0\rangle + |1\rangle)$, the measurement result is random. But the von Neumann entropy of $|+\rangle$ is 0 because it is a pure state.

Just like the Shannon entropy, we can denote the conditional entropy and mutual information of the von Neumann entropy in a similar way. The meanings of them are also similar except the von Neumann entropy is used to measure the uncertainty of a quantum state.

### C. Universal composable security

The universal composable security of a component (a system, a protocol, or a key) is defined as a real component close to an ideal (secure) one. We can say, for example, a key $S$ is $\varepsilon$-secure with respect to an ideal key $K$, which is independent and uniformly distributed, if the distance between $S$ and $K$ is less than or equal to $\varepsilon$, where $\varepsilon$ is a chosen secure parameter that $\varepsilon \geq 0$.

We denote the trace distance between density operators of two quantum states $\rho$ and $\sigma$ by

$$\delta(\rho, \sigma) = \frac{1}{2} tr(|\rho - \sigma|) = \frac{1}{2} tr\left[\sqrt{(\rho - \sigma)^\dagger (\rho - \sigma)}\right].$$

We can say $\rho$ is $\varepsilon$-close to $\sigma$ if $\delta(\rho, \sigma) \leq \varepsilon$. The distance between classical variables can be seen as a special case of the trace distance. Let $X$ and $Y$ be two random variables. The variational distance between the probability distribution of $X$ and $Y$ is denoted by

$$\delta(P_X, P_Y) = \frac{1}{2} \sum_i |P_X(i), P_Y(i)|,$$

where $i$ is the possible value of $X$ and $Y$. The variational distance is equal to the trace distance $\delta([|X|], [|Y|])$, where $[|X|]$ and $[|Y|]$ are the state representations of $X$ and $Y$, i.e.,

$$\delta(P_X, P_Y) = \delta([|X|], [|Y|]).$$

The universal composable security provides many good properties. First, a key that has universal composable security can maintain its security in any context. If a secret key S has universal composable security, the remaining bits in S are secure even some bits of S are given to an adversary. Second, the security of a complex system can be calculated more easily by simply summing the secure parameters $\varepsilon$ of its components. We can do this because the trace distance is subadditive with respect to the tensor product. Third, there are no quantum operators that can increase the trace distance of two quantum states by applying the same operator to them.

## D. Privacy amplification

Let's say that there is a string held by two participants called Alice and Bob, and they want to use it as their secret key. But some information about this string is held by an adversary called Eve, and Alice and Bob only know how many bits have been leaked, though they don't know which ones are leaked. Then, is it possible to use part of this string as a secret key? The answer is affirmative. We can use the privacy amplification to extract a shorter secret key from an insecure string by erasing the information held by Eve. The extracted secret key can have universal composable security and be ε-secure with respect to an ideal key, where ε is the secure parameter chosen by Alice and Bob.

Let $S$ denote the insecure string and $\mathcal{F}$ denote a family of hash functions. The function $\mathcal{F}$ from domain $\mathcal{X}$ to range $\mathcal{Y}$ is said to be two-universal [13, 14] if $\Pr_{f \leftarrow P_\mathcal{F}}[f(x) = f(x')] \leq \frac{1}{|\mathcal{Y}|}$, for any $x, x' \in \mathcal{X}$. The research [10] of the properties of the two-universal hash function shows we can use it to do privacy amplification. The length of the secret key that can be extracted from the insecure string is limited by the information about the string held by Eve. We can bound the key extracted rate (the length of secret key extracted from an insecure string divided by the length of the string) by [15, 16]

$$\text{key extracted rate} \geq S(A|E) - H(A|B) \tag{1}$$
$$= I(A;B) - I(A;E), \tag{2}$$

where $E$ is the quantum information of the string held by Eve and $A$ and $B$ are the string held by Alice and Bob.

In this paper, we use privacy amplification to extract a secret key from a used key. We then call the key extracted rate key recycling rate.

## III. THE PROTOCOL

Before Alice sends a message $Meg \in \{0,1\}^n$ to Bob. Alice and Bob need to share a secret key pool $K$ which they can pick up secret keys from it synchronously. They also share a family of authentication functions $Au()$, an error correction function $ECC_d()$ and a key updating function $Upd()$ defined below.

**Definition 1 [17].** $Au()$ *is a family of hash functions which is ε-almost XOR universal₂ (ε-AXU₂). A family of hash functions $Au_u(Meg) = MAC$ is said to be ε-AXU₂ if for a random key $u$, all $m_1, m_2 \in Meg$ that $m_1 \neq m_2$ and all $t \in MAC$,*

$$\Pr_u[Au_u(m_1) \oplus Au_u(m_2) = t] \leq \varepsilon,$$

*where ε is a secure parameter.*

**Definition 2.** *The error correction function $\text{ECC}_d(Meg) = C$ for $C \in \{0,1\}^{n+s}$ can encode a message $Meg$ into a binary linear code $C$ that has minimum Hamming distance $d$. $ECC_d()$ has a corresponding decoding function that can decode $C$ back to $Meg$.*

When decoding $C$, we can know the errors' number and positions if the error number is not beyond $\frac{d-1}{2}$ and then we can correct $\lfloor \frac{d-1}{2} \rfloor$ errors [18]. If there are too many errors, the decoding of $C$ will fail and output a wrong message.

Before we define the key updating function $Upd()$, we need to define a key recycling function $Rec()$.

**Definition 3.** $Rec(k) = k'$, *where $k$ is the old key, $k'$ is the recycled key and $0 \leq |k'|_l \leq |k|_l$, is a two-universal hash function. We can say $k'$ is ε-secure if for $k_{ideal} \in \{0,1\}^{|k'|_l}$ is an ideal key,*

$$\delta(k_{ideal}, k') \leq \varepsilon.$$

$Rec(k)$ is used to do the privacy amplification to make sure the security of the recycled key. We can then define a key updating function $Upd(k)$ based on $Rec(k)$.

**Definition 4.** *A key updating function $Upd(k) = k'||k^{new} \in \{0,1\}^{|k|_l}$, where $k^{new}$ is a fresh key picked up from $K$ used to extend the length of the output, is used to maintain the security of a key $k$ without shorting its length. We also define the key recycling rate of $Upd(k)$ as $r = \frac{|k'|_l}{|k|_l}$.*

Now we are ready to run our QKR protocol. The steps of our protocol are described below:

1. Alice's classical part: Alice picks up a key: $u \in \{0,1\}^n$ from $K$. Then she computes the message authentication code $MAC = Au_u(Meg) \in \{0,1\}^t$. Alice encodes $Meg||MAC$ into $C = ECC_d(Meg||MAC) \in \{0,1\}^{n+t+s}$. Alice picks up another key: $k_v \in \{0,1\}^{n+t+s}$ from $K$ and computes the cipher $M = C \oplus k_v$.

2. Alice's quantum part: Alice picks up a key: $k_b \in \{0,1\}^{n+t+s}$ from $K$. And encodes $M$ into qubits sequence $Q$ in the z-basis $\{|0\rangle, |1\rangle\}$ or the x-basis $\{|+\rangle = 1/\sqrt{2}(|0\rangle + |1\rangle), |-\rangle = 1/\sqrt{2}(|0\rangle - |1\rangle)\}$ according to $k_b$. Then she sends $Q$ to Bob. After this, Alice recycles $k_b$ and $u$.

3. Bob's quantum part: When Bob receives $Q'$, he picks up keys $u, k_v, k_b$ from $K$ and measures $Q'$ according to $k_b$ and gets the cipher $M' = C' \oplus k_v$.

4. Bob's classical part: Bob computes the plaintext $C' = M' \oplus k_v$. Then he decodes $C'$ to correct potential errors and gets $Meg'||MAC'$, the errors' number $q$ is recorded. Then, Bob computes $MAC = Au_u(Meg')$ and checks $MAC \stackrel{?}{=} MAC'$. Bob updates $k_v$ to $k_v' = Upd(k_v)$ according to $q$ and recycles $k_b$ and $u$.

5. In the next time Bob sends a message to Alice, he will tell Alice $q$ to let Alice update $k_v$ to $k_v'$.

## IV. SECURITY ANALYSIS

In this section, we will analyze the security of the message $Meg$ and the recycled/updated keys $u$, $k_v'$ and $k_b$. These analyses assume the adversary Eve, who wants to get $Meg$ or keys, has full control of the quantum channel between Alice and Bob. Eve also has unlimited computation power and any other resources expect she needs to follow the laws of physics. We will analyze the security under the collective attacks [19] but assume the implementation of the protocol is ideal so Eve cannot use the hacking attacks [20] to get information. When we analyze the security of the recycled keys, we also assume that Eve can do the known-plaintext attack. Since Eve does not know about $u$, she is unable to get $MAC$ in real scenarios. We still assume that Eve can get the whole plaintext (which contains $MAC$) in the analyses of the recycled/updated keys $K_b$ and $k_v'$.

## A. The security of the message Meg

The security of $Meg$ is obvious. After Alice's classical part, the code $C$, which contains the information of $Meg$, is protected by the XOR encryption using $k_v$. As long as $k_v$ is secure, the cipher $M = C \oplus k_v$ has unconditional security and perfect privacy [21].

## B. The security of the recycled u

We first introduce a new family of hash functions Hash() which is ε-almost strong universal$_2$ (ε-ASU$_2$) [22]. A family of hash functions $Hash_u(Meg) = MAC$ is said to be ASU$_2$ if for a random key $u$, all $m_1, m_2 \in Meg$ that $m_1 \neq m_2$ and all $t_1, t_2 \in MAC$,

$$\Pr_u[Hash_u(m_1) = t_1 \text{ and } Hash_u(m_2) = t_2] \leq \frac{\varepsilon}{|MAC|},$$

where $|MAC|$ is the number of $MAC$'s possible values and ε is a secure parameter.

In our protocol, the $MAC$ in the cipher $M$ is protected by $k_v$. Then, $MAC \oplus k_v$ is ε-ASU$_2$ because for secret keys $u$ and $k_v$, all $m_1, m_2 \in Meg$ that $m_1 \neq m_2$ and all $t_1, t_2 \in MAC$,

$$\Pr_{u,k_v}[Au_u(m_1) \oplus k_v = t_1 \text{ and } Au_u(m_2) \oplus k_v = t_2] \leq \frac{\varepsilon}{|MAC|}.$$

Next theorem, derived from the result of earlier research [23], follows:

**Theorem 1**. *We can bound the mutual information of the recycled $u$ and Eve's information EI after $n$ round of the protocol, for $1 \leq n \leq |MAC|$ by*

$$I(u; EI) \leq \log|MAC| - \left(1 - \frac{n}{|MAC|}\right) \log(|MAC| - n),$$

*where EI is the information including $Meg$, $MAC$, and Bob's respond (accept or reject) from these $n$ rounds.*

Theorem 1 shows that the information of $u$ will leak but the leakage is very small and the recycled $u$ is still ε-secure [23].

## C. The security of the message $k_b$

Because our protocol always recycle $k_b$, Eve can just intercept every qubit sent from Alice and don't worry to be detected. After her attack, Eve can also get Bob's response $R$, which telling Eve how many errors Bob detected. The analysis splits into two parts. First, we analyze the information leakage from the intercepted qubits. Second, we analyze the leakage from Bob's response.

When Eve intercepts all qubits sent from Alice, she cannot know the value of each qubit even she can do the known-plaintext attack because the values are protected by $k_v$. Eve needs to distinguish the qubits in the z-basis from the qubits in the x-basis. The next Lemma shows Eve can't get any information about $k_b$ from the intercepted qubits.

**Lemma 1.** Randomly giving Eve one qubit in one of the four states: $|0\rangle, |1\rangle, |+\rangle, |-\rangle$ without giving Eve any other information, Eve cannot know the basis of the qubit.

*Proof.* We assume Eve can measure the qubit with any ancilla bits and use any measurement operators. And we denote the qubit with the ancilla bits as $|\varphi_i\rangle$, for $i \in \{0, 1, +, -\}$, and measurement operators are $\{M_m\}$. Base on the postulates of quantum measurement [24], the probability of measurement outcome $m$ of the state $|\varphi_i\rangle$ is $p(m)_i = \langle \varphi_i | M_m^\dagger M_m | \varphi_i \rangle$, where $M_m^\dagger$ is the conjugate transpose of $M_m$. By direct computation, for all $M_m$ and all corresponding $m$,

$$p(m)_0 + p(m)_1 = p(m)_+ + p(m)_-.$$

When Eve gets the measurement outcome $m$, she cannot get any information about the basis of the qubit. □

Next, we analyze the information leakage from Bob's response. After Bob receiving the qubits from Alice (or Eve), he uses ECC to detect and correct the potential errors and sends the result back to Alice in the next round of the protocol. The response $R$ is an integer which $0 \leq R \leq |K_b|_l$. We assume Eve can know the response $R$ and use it to get $k_b$. For example: Eve can measure a qubit in the z-basis, and then wait for the response by Bob. If Bob detects one error, Eve will know that her measurement caused this error and that the basis of the qubit is not in the z-basis.

According to the analysis above, the information leakage is only from Bob' response $R$. By the properties of mutual information [25] that the mutual information $I(k_b; E')$ will not increase more than the exchanging information between the participants and Eve, We can bound the increasing of $I(k_b; E')$ by the Shannon entropy of $R$. Next theorem gives the desired result.

**Theorem 2.** *We can bound $I(k_b; E')$, where $E'$ is Eve's information after this round of the protocol, by*

$$I(k_b; E') \leq I(k_b; E) + H(R),$$

*where $I(k_b; E)$ is the mutual information of recycled $k_b$ and Eve's information before this round of the protocol.*

We can cover the information leakage from R by slightly updating the $K_b$ using $Upd()$. The key recycling rate of $Upd(K_b)$ is bound by Equation (2) and is nearly 1 when the length of the message is long enough.

## D. The security of the message $k'_v$

The security of $k'_v$ is easy to achieve. For example, $Rec(k_v)$ can just output nothing and $k'_v$ is a new fresh key picked from $K$. In this section, we focus on how many secret keys can be recycled from the old $k_v$ at most, i.e., the optimal key recycling rate of $Udp(k_v)$. To calculate the key recycling rate of $Udp(k_v)$, we use the properties of privacy amplification and bound the key recycling rate by Equation (1). We describe the formula here again:

$$r \geq S(A|E) - H(A|B),$$

where $r$ is the key recycling rate, $A$ and $B$ are $K_v$ held by Alice and Bob respectively and $E$ is the quantum state held by Eve. [15] shows the information leakage from the error correction part is $H(A|B)$. Since in our QKR protocol, Bob already knows $K_v$. They don't need to synchronize the key after the protocol and will not leak any information from the error correction part. The next theorem shows the effect of this difference.

**Theorem 3.** *We can get $H(A|B) = 0$ because Alice and Bob already know $k_v$. The key recycling rate of $Udp(k_v)$ is then only bounded by*

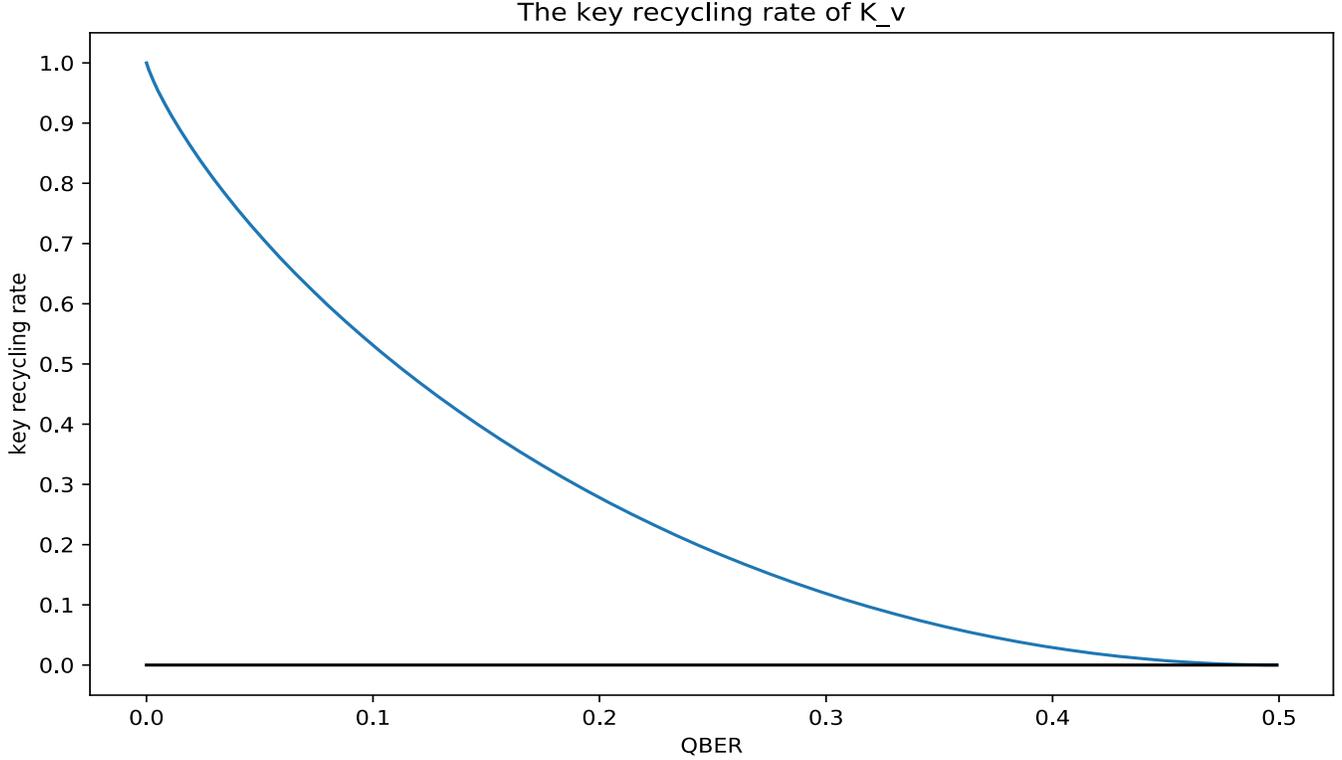

Fig. 1. The x-axis denotes the real QBER and the y-axis denotes the key recycling rate of $Upd(K_v)$.

$$r \geq S(A|E). \quad (3)$$

Now we can calculate the key recycling rate of $Udp(k_v)$. We denote the joint system held by Alice, Bob and Eve after Eve's attack by

$$|\psi\rangle_{ABE} := \sum_{i=1}^{4} \sqrt{\lambda_i} |\Phi_i\rangle_{AB} \otimes |v_i\rangle_E ,$$

where $|\Phi_1\rangle_{AB}, \dots, |\Phi_4\rangle_{AB}$ denote four Bell states in Alice and Bob's joint system and $|v_1\rangle_E, \dots, |v_4\rangle_E$ are some mutually orthogonal states held by Eve. Depending on Alice and Bob's measurement results (with respect to the z-basis), we can denote Eve's system by $|\theta^{a,b}\rangle$, where $a, b$ are Alice and Bob's measurement results,

$$|\theta^{0,0}\rangle = \frac{1}{\sqrt{2}}(\sqrt{\lambda_1}|v_1\rangle_E + \sqrt{\lambda_2}|v_2\rangle_E)$$

$$|\theta^{0,1}\rangle = \frac{1}{\sqrt{2}}(\sqrt{\lambda_1}|v_1\rangle_E - \sqrt{\lambda_2}|v_2\rangle_E)$$

$$|\theta^{1,0}\rangle = \frac{1}{\sqrt{2}}(\sqrt{\lambda_3}|v_3\rangle_E + \sqrt{\lambda_4}|v_4\rangle_E)$$

$$|\theta^{1,1}\rangle = \frac{1}{\sqrt{2}}(\sqrt{\lambda_3}|v_3\rangle_E - \sqrt{\lambda_4}|v_4\rangle_E).$$

We can write the density operator of Eve's state according to Alice's measurement results with respect to the basis $\{|v_1\rangle_E, \dots, |v_4\rangle_E\}$,

$$\sigma_E^a = \begin{bmatrix} \lambda_1 & \pm\sqrt{\lambda_1 \lambda_2} & 0 & 0 \\ \pm\sqrt{\lambda_1 \lambda_2} & \lambda_2 & 0 & 0 \\ 0 & 0 & \lambda_3 & \pm\sqrt{\lambda_3 \lambda_4} \\ 0 & 0 & \pm\sqrt{\lambda_3 \lambda_4} & \lambda_4 \end{bmatrix},$$

where $\pm$ is a plus sign if $a = 0$ and a minus sign if $a = 1$. Now we are ready to calculate Equation (1). Using the fact that $S(AE) = H(A) + S(E|A)$, we can get

$$S(A|E) = S(E|A) + H(A) - S(E)$$

with

$$S(E|A) = \frac{1}{2} S(\sigma_E^0) + \frac{1}{2} S(\sigma_E^1),$$

$$S(E) = S(\sigma_E^0 + \sigma_E^1),$$

and $H(A) = 1$. Our protocol uses two encodings: the z-basis and the x-basis. Bob can know the Quantum Bit Error Rate (QBER) for both encodings by the ECC. This gives two conditions on the state held by Alice and Bob. If we use the Bell basis to express these conditions, and if the QBER equals $Q$ for both encodings, we can get $\lambda_3 + \lambda_4 = Q$ and $\lambda_2 + \lambda_4 = Q$, where $\lambda_1, \dots, \lambda_4$ are the diagonal entries of Alice and Bob's states (with respect to the Bell basis). A straightforward calculation shows when $\lambda_4 = Q^2$, the key recycling rate takes its minimum. We show the result in Fig. 1.

The key recycling rate shows in Fig. 1 is the rate when the authentication check is successful. If the authentication check fails, the key recycling rate is just 0. This is because in our protocol, we use the ECC and $MAC$ to get the quantum bit error rate (QBER) instead of using the classical authentication channel to do the public discussion. When the authentication check succeeds, Bob has a $1-\varepsilon$ chance to consider the message is from Alice and is not disturbed by Definition 1. We can then consider the errors that were corrected by the ECC from $K_v$. We can know how many errors there are when the number of errors is not higher than $\left\lfloor \frac{d-1}{2} \right\rfloor$ by the ECC. When the error number is higher than $\left\lfloor \frac{d-1}{2} \right\rfloor$, the decoding will fail and output a wrong message, and the authentication check will fail, too. If the authentication check fails, we need to set the QBER to 50% to cover the worst case because the ECC cannot output the right message and the QBER.

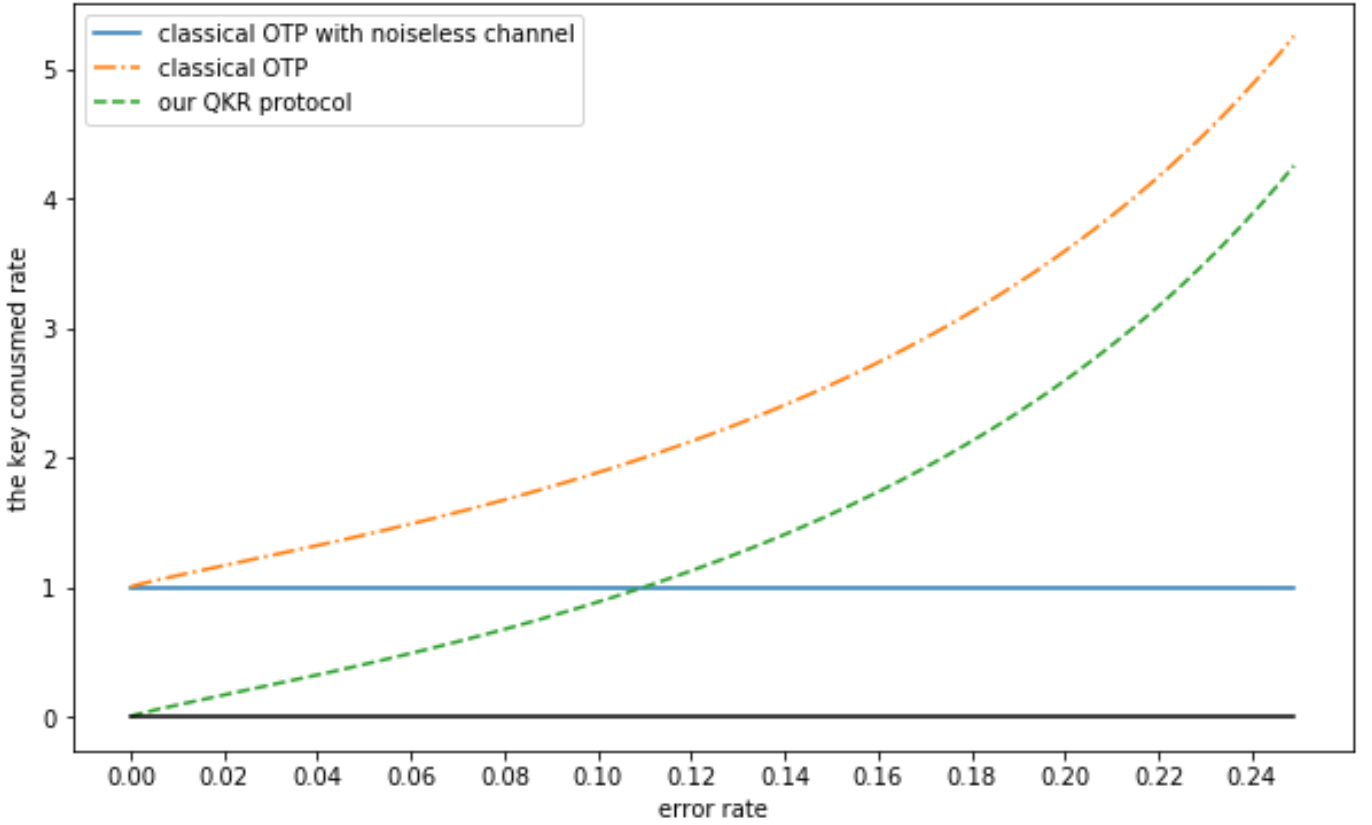

Fig. 2. The x-axis denotes the error rate $Q$, and the y-axis denotes the key consumed rate (the bit numer of consumed pre-shared key / the bit number of message). The line classical OTP with noiseless channel is always 1 because in a noiseless channel, the message does not need ECC. The line classical OTP shows the key consumed rate of a classical OTP with a noise channel, which has error rate $Q$. And the line our QKR protocol shows the lowest key consumed rate we can achieve according to $Q$ in our QKR protocol.

## V. THE EFFICIENCY OF OUR QKR PROTOCOL

It is obvious that our QKR protocol is more efficient than a classical OTP, which cannot recycle any used keys. Our QKR protocol can recycle part of $K_v$ and almost all other keys. When the QBER is 50%, we can consider the whole quantum message is intercepted by Eve and replaced by a random quantum sequence, and this is the worst situation. Even we need to abandon the entire $K_v$ in this worst case, the amount of keys we spend to send a message is equal to the classical OTP.

To show more detail of the efficiency of our QKR protocol, we will calculate how many pre-shared keys are consumed in our QKR protocol and compare the result to the classical OTP (as shown in Fig. 2). The real number of consumed keys in our QKR protocol is not only according to the key recycling rate but also according to the length of the used pre-shared keys. The length of used key $K_v$ equals the length of the code $C$, which is

$$\left(1 + \frac{H(Q_p)}{1-H(Q_p)}\right) * n,$$

where $Q_p$ is the prediction error rate of the quantum channel, and $n$ is the length of the message $Meg$. We ignore the consumed keys $u$ and $K_b$ because the key recycling rate of $u$ and $K_b$ is almost 1 when the message is long enough. Combine the above. The total consumed key of our QKR protocol can be denoted by

$$\left(1 + \frac{H(Q_p)}{1-H(Q_p)}\right) * n * (1 - \text{the key recycling rate of } K_v).$$

Fig. 2 shows our QKR protocol can use less key to send a secret message than a classical OTP with the noiseless channel when the quantum channel's error rate is less than 0.11. And when the error rate of the classical channel and the error rate of the quantum channel are the same, our QKR protocol can always send the secret key with less consumed key (when the error rate is 0.5, the key consumed rate of classical OTP and the key consumed rate our QKR protocol are the same).

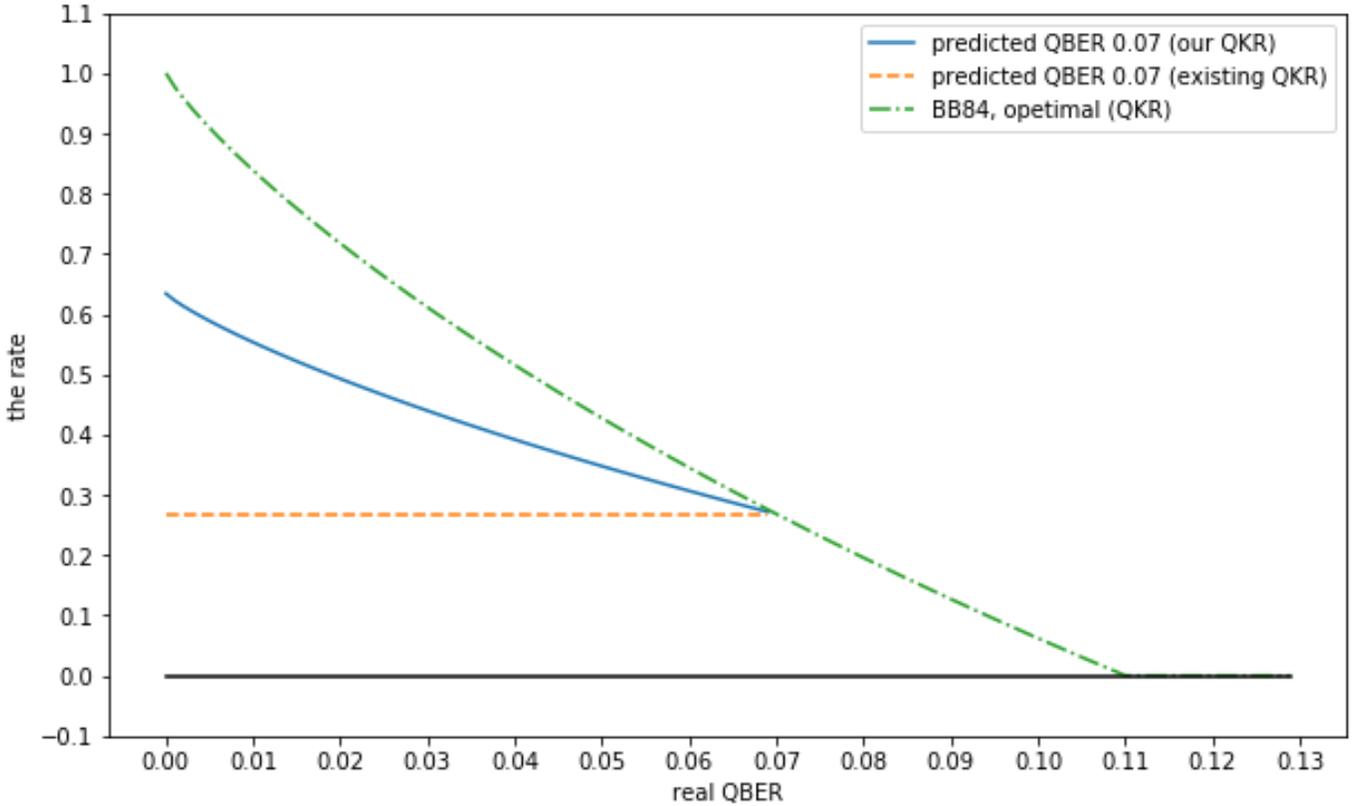

Fig. 3. The x-axis denotes the real QBER $Q$, and the y-axis denotes the rate. The line predicted QBER 0.07 (our protocol) shows the rate of our QKR protocol with the predicted QBER $Q_{predict} = 0.07$ according to $Q$. When $Q = 0.07$, this is the best rate out of all $Q_{predict}$ for our QKR protocol. The line predicted QBER 0.07 (existing QKR) shows the rate of existing QKR with the prediction QBER $Q_{predict} = 0.07$. And the line BB84 shows the key rate of the BB84 protocol according to $Q$. This is also the best rate of QKR protocol we can achieve according to $Q$ (i.e., the rate when $Q_{predict} = Q$)

We also compared our QKR protocol with the existing QKR protocols with conjugate coding [4, 6], which the rates are the same with the key rate of BB84. The rate[2] is denoted as

$$\frac{|message|_l - |consumed\ key|_l}{the\ number\ of\ all\ qubits}.$$

These protocols use part of the message to share new keys to cover the consumed keys. To compare our QKR protocol with theirs, we assume our QKR protocol using part of the message to share new keys but not picking keys from the key pool to cover the consumed keys (though our QKR protocol cannot predict how many keys will be consumed). The result shows our QKR protocol is more efficient than theirs when the real QBER is less than the predicted QBER (as shown in Fig. 3). It is worth mention that our QKR can send messages even when the error rate is beyond 0.11 and the rate is negative because the consumed key is actually covered by the fresh key picked from the key pool in our QKR protocol.

## VI. CONCLUSION AND OPEN QUESTIONS

In this paper, we propose a new QKR protocol which can recycle used pre-shared key according to the real error rate in the quantum channel. Our QKR protocol has higher efficiency than the existing QKR protocols, and our QKR protocol can run with a more noisy channel. By delaying recycling some of the keys, the participants do not need any classical authentication channel in our QKR protocol. Finally, the universal composable of the recycled key is proved by using the information theory.

In this study, our analysis is under the condition that there is an optimal ECC, and the length of transmission messages is infinite. However, the analysis with finite resources can help us understand the potential of the QKR protocols in the practical situation. Therefore, how to analyze the security of our QKR protocol with finite resources [26, 27] will be our future work. Additionally, the core idea that uses the error rate to optimize the key recycling rate in our research may also be used to improve the key recycling rate of quantum authentication protocol [28], which can only recycle the used keys when the quantum channel is noiseless. However, the security proof stated in this paper cannot be directly used in the quantum authentication protocol. The complete and suitable security analyses for key recycling rate optimization in quantum authentication protocol are also important future work.

---

[2] The definition of the rate of QKR protocols uses the number of transmitted qubits to be the denominator. But the key rate of QKD protocols use the number of the raw keys (in regular BB84, the number of the raw keys is merely 1/4 of the number of all transmitted qubits) to be the denominator. Because our QKR protocol uses some qubits to send the redundant information added by error correcting function, the rate of our QKR protocol can not achieve 1 even when the real error rate is 0. But we will not loss any pre-shared key when the real error rate is 0 actually.

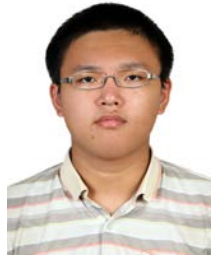

**Yu-Chin Lu** was born in Kaohsiung, R.O.C. in 1995. He received the M.S. degrees in computer science and information engineering from the National Cheng Kung University, Tainan, in 2020.

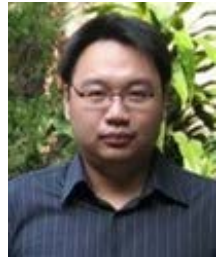

**Chia-Wei Tsai** received the Ph.D. degrees in computer science and information engineering from the National Cheng Kung University, Taiwan, in 2011. He is currently a Assistant Professor with the Department of Computer Science and Information Engineering, National Taitung University, Taitung, Taiwan. He has authored over 45 technical papers and holds two patents. His research interests include quantum information and machine learning.

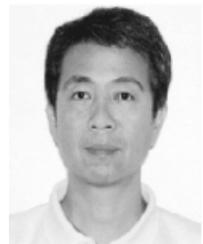

**Tzonelih Hwang** received the M.S. and Ph.D. degrees in computer science from the University of Southwestern Louisiana, USA, in 1988. He is currently a Distinguished Professor with the Department of Computer Science and Information Engineering, National Cheng Kung University, Tainan, Taiwan. He has actively participated in several research activities, including a Research Scientist with the Center for Advanced Computer Studies, University of Southwestern Louisiana. He is also a Member of the Editorial Board of some reputable international journals. He has authored over 250 technical papers and holds five patents. His research interests include network and information security, access control systems, error control codes, security in mobile communication, and quantum cryptography.